\documentclass[aip, jap, reprint, floatfix]{revtex4-1}

\usepackage{graphicx}
\usepackage{amsmath}
\usepackage{color}
\usepackage{soul}

\DeclareMathOperator{\sign}{sign}

\newcommand{\A}{{\rm \AA}}
\newcommand{\diff}{\mathrm{d}}
\newcommand{\bk}{\mathbf{k}}
\newcommand{\br}{\mathbf{r}}
\newcommand{\gt}{>}
\newcommand{\dFD}{\frac{\partial f_{\rm FD}(\epsilon)}{\partial \epsilon}}
\newcommand{\Oslo}{Centre for Materials Science and Nanotechnology (SMN), University of Oslo, P.O.B. 1126 Blindern, NO-0318 OSLO, Norway}
\newcommand{\OsloB}{Department of Physics, University of Oslo, P.O.B. 1048 Blindern, NO-0316 OSLO, Norway}
\newcommand{\Sintef}{SINTEF Materials and Chemistry, Forskningsveien 1, NO-0314 OSLO, Norway}

\begin{document}

\title{Enhancement of thermoelectric properties by energy filtering: Theoretical potential and experimental reality in nanostructured ZnSb}
\author{Kristian Berland}    \affiliation{\Oslo}
\author{Xin Song}     \affiliation{\OsloB}
\author{Patricia A. Carvalho}\affiliation{\Sintef}
\author{Clas Persson} \affiliation{\Oslo}\affiliation{\OsloB} 
\author{Terje G. Finstad}  \affiliation{\Oslo}\affiliation{\OsloB}
\author{Ole Martin L\o vvik}   \affiliation{\OsloB}\affiliation{\Sintef}

\begin{abstract}
Energy filtering has been suggested by many authors as a means to improve thermoelectric properties. The idea is to filter away low-energy charge carriers in order to increase Seebeck coefficient without compromising electronic conductivity. This concept was investigated in the present paper for a specific material (ZnSb) by a combination of first-principles atomic-scale calculations, Boltzmann transport theory, and experimental studies of the same system. The potential of filtering in this material was first quantified, and it was as an example found that the power factor could be enhanced by an order of magnitude when the filter barrier height was 0.5~eV. Measured values of the Hall carrier concentration in bulk ZnSb were then used to calibrate the transport calculations, and nanostructured ZnSb with average grain size around 70~nm was processed to achieve filtering as suggested previously in the literature. Various scattering mechanisms were employed in the transport calculations and compared with the measured transport properties in nanostructured ZnSb as a function of temperature. Reasonable correspondence between theory and experiment could be achieved when a combination of constant lifetime scattering and energy filtering with a 0.25~eV barrier was employed. However, the difference between bulk and nanostructured samples was not sufficient to justify the introduction of an energy filtering mechanism. 
The reasons for this and possibilities to achieve filtering were discussed in the paper.
\end{abstract}

\maketitle

\section{Introduction}
 Thermoelectric materials allow for the conversion of temperature gradients to electricity and vice versa. They are today mainly used within sectors such as automotive, aerospace, defense, industrial and self-powered sensors.  For direct power generation the low efficiency is the major technical factor limiting the growth of the market.\cite{Telkes1947, Snyder2008ECS, Zebarjadi20125147, Radousky2012502001}

Good thermoelectric materials are distinguished by 
 low thermal conductivity $\kappa$, high electronic conductivity $\sigma$ and
high Seebeck coefficient ($S$) at a given temperature $T$. 
This can be quantified by the dimensionless 
figure of merit $ZT$
\begin{align}
  ZT =  \frac{ \sigma S^2 T }{\kappa_e +\kappa_l}.  
  \label{eq:ZT}
\end{align}
Due to the Wiedemann-Franz law linking $\sigma$ closely together with the electron part of the thermal conductivity $\kappa_e$,\cite{chasmar195952}
much emphasis is put on lowering the lattice thermal conductivity $\kappa_l$. The power factor $PF=\sigma S^2$ should furthermore be maximized by choosing the optimal charge carrier concentration.
We have in this paper demonstrated that this last requirement entails electronic conditions favoring transport of high-energy over low-energy carriers.

Nanostructured materials offer new mechanisms to selectively scatter phonons and low-energetic electrons without strongly affecting the transport of energetic electrons.  \cite{Zebarjadi20125147, Faleev2008, Narducci2015} Efficient bulk thermoelectric materials are a good starting point for further nano-enhancements; yet even poor ones may serve  --- nanostructured silicon have for example
shown promising thermoelectric properties.\cite{Bux20092445}  A particularly interesting concept is that of energy-filtering. By introducing potential barriers or strongly energy-dependent scattering mechanisms low-energetic carriers can be blocked, greatly enhancing the Seebeck coefficient.\cite{EnergyFiltering:superlattices,EnergyFiltering:GrainBoundaries,Dehkordi:EnergyFiltering,Nolas:Filtering,Bahk2013:filtering, Yang2013, Narducci201219,flage-larsen2012,Popescu2009205302, Zou2014,Narducci2015}

ZnSb has been known as a thermoelectric material for a long time.\cite{Shaver1966} When Caillat reported a figure of merit of 1.4 for Zn$_4$Sb$_3$ in 1997 that composition got the most attention due to the remarkably low thermal conductivity.\cite{Snyder2004458} ZnSb was then mostly regarded as an annoying phase impurity.   However, two phase transitions, one from the $\alpha$ to $\beta$ phase at 250~K, and one from $\beta$ to $\gamma$ at 767~K,\cite{Izard2001567} make Zn$_4$Sb$_3$ difficult to use in applications.  ZnSb has received renewed interest \cite{Xiong2013397,Fedorov2014JEM} for a number of reasons. There is an increased awareness of environmental concerns, where Zn and Sb score well for abundancy and low toxicity. There is also a lack of other good alternative materials for operation in the temperature-range $400-650$~K, where ZnSb performs well. Further, the thermoelectric properties of bulk ZnSb are suitable for improvement by nanostructuring.\cite{Vineis2010,Dresselhaus2007} Several reports on densely packed pellets of ZnSb have appeared recently, utilizing techniques like ball-milling,\cite{Bottger20112753, Kjetil2012, Bottger2010, Okamura2010, Niedziolka2014} spark plasma sintering,\cite{Blichfeld2015} and cryogenic milling.\cite{xins20152578} Optimization of doping levels and alloying elements have significantly enhanced the thermoelectric properties of ZnSb,\cite{Kjetil2012, Bottger2011, Xiong2013397, Fedorov2014semi} utilizing the potential of the impurity band.\cite{xins2012, Bottger2011, Valset2015, Eklof2013}  This has led to an improvement of the figure-of-merit from 0.3 in the 1960's \cite{Justi196427} to consistent reports of $zT\gt0.9$.\cite{Kjetil2012, Xiong2013397, Fedorov2014JEM}

A number of theoretical studies of ZnSb have been reported in recent years. {\em Ab initio} band structure calculations have been reported by several groups.\cite{Mikhaylushkin2005, Bottger2011, Benson2011, Zhao2011, Bjerg2011, Bjerg20122111, Jund201285, Niedziolka2014, Haussermann2010}  These have e.g.\ allowed comparisons with experimental effective masses,\cite{Bottger2011}  the stability of the material,\cite{mozharivskyj2004} vacancy formation energies revealing the nature of the bonding \cite{Jund201285} and charge transfer to bonds or neighbor atoms.\cite{Niedziolka2014, Niedziolka2015}   Also, a few phonon dispersion results and studies addressing thermal properties of ZnSb from first principles have recently appeared.\cite{Jund201285, Bjerg2014, Fischer224309, Hermet87118}

\begin{figure*}[t!]
\includegraphics[width=1.95\columnwidth]{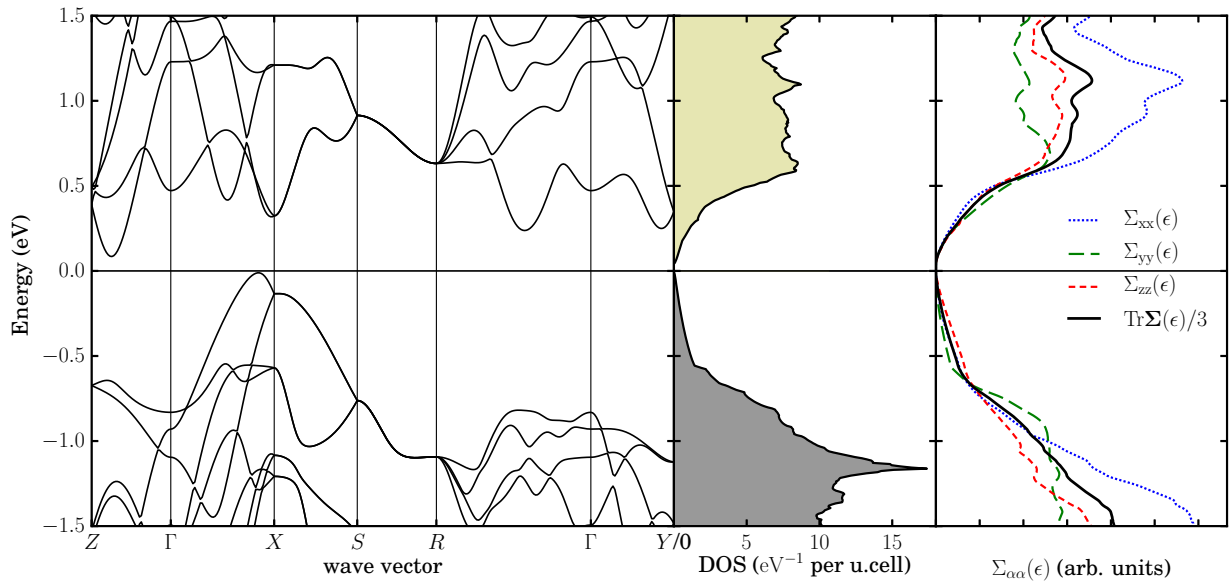}
\caption{Electronic band structure of ZnSb obtained using the Perdew-Burke-Ernzerhof (PBE) exchange-correlation functional (left), corresponding density of states (DOS) (middle), diagonal elements of the tensorial transport spectral function  (right). See text for explanation. \label{fig:electonicStruc} }
\end{figure*}

In this paper, we have quantified the theoretical potential of energy filtering in ZnSb, demonstrating that it is indeed possible, from a theoretical point of view, to greatly enhance the power factor of ZnSb. 
In an attempt to verify this experimentally, 
we prepared nanostructured ZnSb samples by a combination of cryomilling and rapid hot pressing, which has been shown previously to generate densely packed pellets with very small grain size and significantly reduced thermal conductivity.\cite{xins20152578}
Our hypothesis was that such processing could introduce energy filtering from grain boundaries or nanoinclusions associated with grain boundaries.\cite{Narducci201219,Popescu2009205302,Bahk2013:filtering}
The transport properties of these samples were then compared to the theoretical predictions with and without energy filtering.

This paper is organized as follows:
First,  a brief description of the sample preparation and experimental methods are provided. 
Then follows the theoretical approach to calculating thermoelectric properties and solving
the Boltzmann transport equation with different scattering models including energy filtering. 
This is followed by an analysis of the potential of energy filtering of ZnSb.
A comparison between theory and experiment for a bulk reference sample is then presented,
validating the approach qualitatively and indicating quantitative shortcomings.
The final part is a 
comparison between theory and experiment for nanostructured samples.

\section{Experimental methods}\label{sec:experimental-methods}
Starting from stoichiometric Zn and Sb sealed in evacuated quartz tube without any intentional dopants, the initial materials were synthesized by melting and solidification. The mix was melted at 970~K followed by quenching in cold water. 
Two thermo-mechanical processing routes were then followed: (i) a "nanostructured" sample was produced by ball milling at 77~K and hot-pressing at 740~K for 30 min and cooling to RT within 2 h; (ii) a "bulk" sample, used as reference, was produced by ball milling at room temperature and hot-pressing at 740~K for 30 min and cooling to RT within more than 20 h.
Further details on the fabrication method, reduction of thermal conductivity in nanostructured samples, etc.\ are described in Ref~\onlinecite{xins20152578}.

A number of different methods were used to characterize the samples:
The microstructure was investigated by transmission electron microscopy and energy dispersive spectroscopy (EDS) using an FEI Titan G2 60-300 instrument operated at 300 kV. For better statistics, the average grain size was estimated from the full-width half maximum (FWHM) of X-ray diffraction peaks using the TOPAS software, which includes information about the instrument contributions in the peak shape analysis.\cite{topas} 
The Seebeck coefficient was measured with the uniaxial four-point method in vacuum.\cite{Iwanaga2011}
  Finally, the electrical conductivity and carrier concentration were measured in vacuum with the Van der Pauw and Hall methods using a custom-built instrument.\cite{Borup2012}

  \section{Theory}

  The Boltzmann transport equation 
  in the relaxation-time approximation 
  was used  to calculate thermoelectric properties.
  As input for these calculations, we used the electronic band structure from density functional theory calculations together with a specified energy filtering and constant relaxation time $\tau$.
  These results were also compared with results obtained with a 
  simple energy ($\epsilon$)-dependent scattering of the form 
\begin{align}
\tau(\epsilon) = \tau_s \left( \epsilon/k_{\rm B} T\right)^s, 
\label{eq:tau}
\end{align}
where the scattering parameter $s$ determines the energy dependency and thus the specific scattering mechanism. $k_{\rm B}$ is the Boltzmann constant. Important examples include acoustic-phonon scattering ($s=-0.5$), polar optical phonon scattering ($s=0.5$), and ionized impurity scattering ($s=1.5$).\cite{Pichanusakorn201019} The net effect of less energy-dependent scattering mechanisms, such as scattering from neutral defects, can be represented by a constant lifetime contribution ($s=0$). The various possibilities represented by equation (\ref{eq:tau}) can account reasonably well for typical scattering mechanisms existing in bulk materials, at least for scattering around nondegenerate band minima.\cite{lundstrom2009fundamentals} 
   
   Energy filtering was implemented in these calculations by simply 
  removing contributions to the thermoelectric transport properties that arise from charge carriers close to the valence band edge. 
  According to theoretical considerations, energy filtering can arise from extended barriers
  such as heterostructures, nanocomposites, nanoinclusions, or grain boundaries.\cite{EnergyFiltering:superlattices,EnergyFiltering:GrainBoundaries,Dehkordi:EnergyFiltering,Nolas:Filtering,Bahk2013:filtering,Narducci201219,flage-larsen2012,Popescu2009205302}

 \subsection{Electronic structure calculations}

The structure and electronic properties of ZnSb were calculated utilizing the plane wave code \textsc{VASP}, working at the density functional theory (DFT) level and using the projector augmented wave approximation for atomic core regions.\cite{vasp1,vasp2,vasp3,vasp4} The generalized gradient PBE~\cite{pebuer96} exchange-correlation functional was used and spin-orbit coupling was ignored. 

  To obtain the atomic and crystal structure, we relaxed the structure with DFT with an energy cutoff of 500~eV, which is 80\% larger than the standard recommended maximum pseudopotential cutoff. Such high cutoffs are needed to accurately determine the structure. 
The $\bk$-point sampling was set to $10\times8\times8$ and due to the low PBE band gap, the Gaussian smearing was set to 0.03 eV. 
The structure was relaxed until forces became smaller than 0.02~eV/\A. 
The calculated lattice parameters of the orthorhombic unit cell,
6.28~\A, 7.82~\A, and 8.22~\A, agree well with previous calculations.\cite{materialsproject,materialexplorer} 
For comparison, the experimental values at room temperature are 6.218~\A, 7.741~\A, and 8.115~\A.\cite{Almin1948}

To obtain the electronic structure, we first generated the electronic charge density $n(\br)$, using an energy cutoff of 276~eV, corresponding to the recommended maximum pseudopotential cutoff and a dense $\bk$-mesh of $20\times16\times16$ integrated using the  tetrahedron method with Bl\"ochl corrections. The total energy was converged to $10^{-6}$ eV.
  Following this step, we generated the band structure with a non-selfconsistent DFT calculations with a  $\bk$-mesh of $50\times50\times50$, as such very dense meshes are required for accurate transport properties.

  Figure \ref{fig:electonicStruc} shows the electronic band structure of ZnSb~(left), density of states $\rho(\epsilon)$~(middle), and diagonal elements of the tensorial transport spectral functions $\bf{\Sigma}(\epsilon)$\cite{Mahan1996_7436} for a constant relaxation time (right). 
  $\bf{\Sigma}(\epsilon)$ is a $3\times3$ tensor, and its diagonal elements are defined in the following manner:
  \begin{align}
    \Sigma_{\alpha\alpha}(\epsilon) &=\frac{1}{V N} \sum_{\bk,i}\,(\nu_\alpha^i(\bk) )^2\tau_i(\bk) \, \delta\left( \epsilon -\epsilon^i(\bk) \right)\,,
    \label{eq:Sigma}\,
  \end{align}
  where $V$ is the volume, $N$ is the number of majority charge carriers, $\tau_i(\bk)$ is the relaxation time for band number $i$, $\epsilon^i(\bk)$ is the energy of band $i$ at reciprocal vector $\bk$, and $\nu_\alpha^i(\bk)$ is the group velocity in the $\alpha$-direction ($\alpha=x, y, z$).
   
The most relevant region for low-field transport is that close to the band edges (for energies less than e.g.\ 0.5~eV away from the Fermi level). We first note that, close to the band edges, the level of anisotropy for $\bf{\Sigma}(\epsilon)$ is somewhat higher (the relative difference between the diagonal components is larger) in the valence-band region than in the conduction-band region. We have in the remainder of the paper neglected the anisotropy by assuming that the samples are multicrystalline and isotropic on average. This was imposed by using the mean of the diagonal elements of the transport spectral function:  $\Sigma(\epsilon) ={\rm Tr}\left( {\bf \Sigma}(\epsilon)\right)/3$. The spectral functions are on the other hand larger in magnitude above the conduction band minimum (CBM) than below the valence band maximum (VBM). This can be rationalized from the shape of the band structure (left) having a single dominant peak near the VBM and multiple ones of relatively similar energy near the CBM.

  The presence of an impurity band originating from Zn defects can explain many of the features of ZnSb at low temperatures, and a model involving single parabolic bands including an explicit impurity band was rather successful in reproducing transport properties of intentionally undoped ZnSb.\cite{xins2012} In the present study, we have chosen to include contributions from such impurities as an effective scattering model combined with adapting the charge carrier concentration by changing the Fermi level. The alternative, introducing an explicit impurity band to the calculated band structure as in Ref.~\onlinecite{xins2012}, would imply ambiguities related to the position and size of the impurity band. One could include the impurity band indirectly by adding Zn vacancies (the most stable intrinsic impurity in ZnSb) as in Ref.\ \onlinecite{Jund201285}, but this would make it difficult to fine-tune the doping level, particularly without involving prohibitively large supercells. Also, our choice gave the ability to directly compare contributions from impurity scattering with other mechanisms.

  \subsection{Boltzmann transport equation}

\begin{figure*}
    \includegraphics[width=0.95\columnwidth]{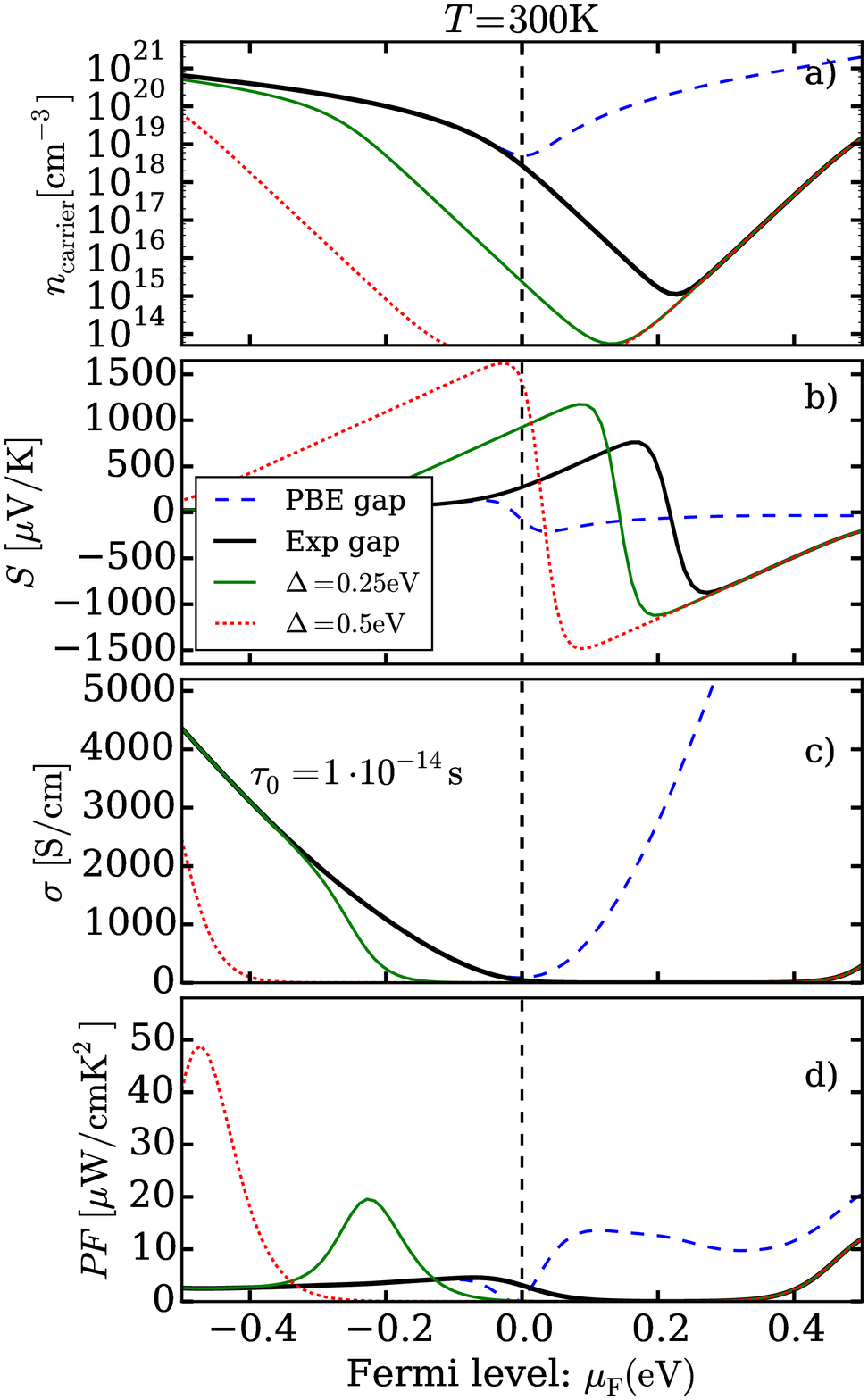}
    \includegraphics[width=0.95\columnwidth]{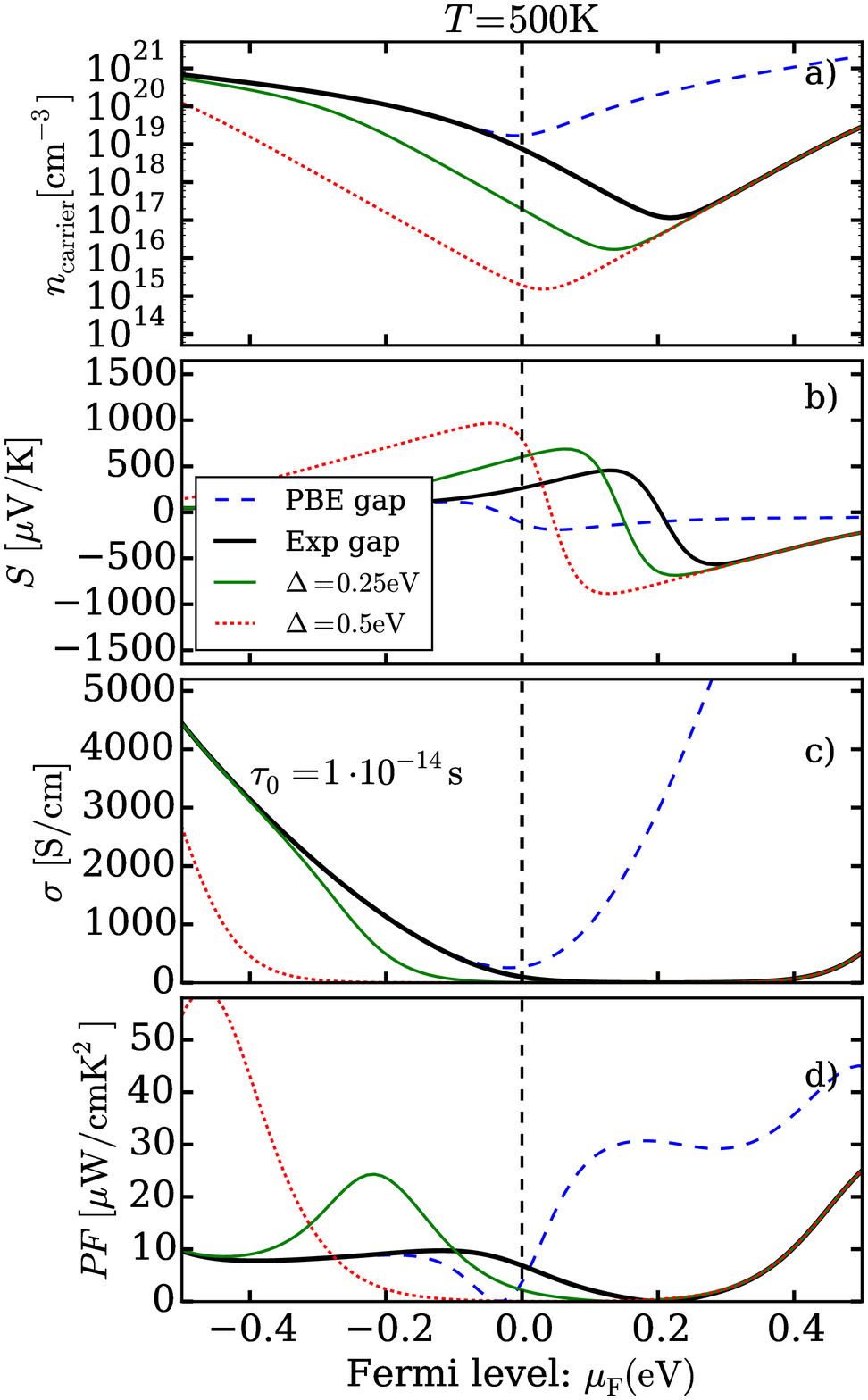}
    \caption{Calculated (Hall) carrier concentration (a), Seebeck coefficient (b), conductivity (c), and power factor (d) of ZnSb at $T=300\,{\rm K}$ (left panels) and  $T=500\,{\rm K}$ (right panels) as a function of the Fermi level $\mu_{\rm F}$. A constant scattering time with $\tau_0=1\cdot10^{-14}$~s was used. The dashed blue curves are based on the PBE band gap (0.06~eV), while the thick black curves rely on the experimental band gap (0.56~eV). The thin green and dotted red curves are results with a valence band energy filter of $0.25~\rm{eV}$ and  $0.5~\rm{eV}$. \label{fig:Thermo_mu}  }
  \end{figure*}

Key thermoelectric quantities can be expressed in terms of integrals of the transport-spectral function $\Sigma(\epsilon)$ as follows
  \begin{align}
    \sigma &=  e^2 \int_{-\infty}^\infty \diff \epsilon\, \left(-  \dFD \right) \Sigma(\epsilon) \nonumber\,, \\
    T\sigma S &= e  \int_{-\infty}^\infty \diff \epsilon\,\left( - \dFD \right)\Sigma(\epsilon) (\epsilon-\mu_{\rm F})\nonumber \,,\\
    T\kappa_0 &=   \int_{-\infty}^\infty \diff \epsilon\,\left( - \dFD \right)\Sigma(\epsilon) (\epsilon-\mu_{\rm F})^2 \,.\label{eq:thermo}
  \end{align}

 Here the derivative of the Fermi-Dirac distribution function $ \left( - \dFD \right)$ is the Fermi window, a symmetric function peaked when the energy $\epsilon$ is equal to the Fermi level, $\mu_{\rm F}$. 

Our calculated PBE band gap of ZnSb was $0.06~{\rm eV}$, which is consistent with previous studies at the same level of theory.\cite{Niedziolka2014, Jund201285, Benson2011} This level of theory is known to severely underestimate the gap compared to  experimental values.  The typical experimental value of the band gap for single crystal ZnSb is $0.5-0.6$~eV \cite{Turner1961, Shaver1966, ref:bandgap, Haussermann2010}. However, there are also experimental reports of a ZnSb band gap around 0.3~eV.\cite{ Zhang2003} We chose to enlarge the calculated band gap by 0.5~eV in order to be consistent with recent {\em ab initio} studies employing the more reliable Heyd-Scuseria-Ernzerhof (HSE) hybrid functional, where the band gap was predicted to be 0.56~eV.\cite{Niedziolka2014, Niedziolka2015} The adjustment was implemented by a simple scissor operator widening the gap in $\Sigma(\epsilon)$ and $\rho_i(\epsilon)$ by 0.5~eV, keeping their shapes otherwise fixed.

Energy filtering corresponding to a nonplanar potential \cite{Bahk2013:filtering} was implemented by removing the contributions from the top of the valence band region ($\epsilon=0$) in a width $\Delta$, as expressed in terms of Heaviside step functions $h$ as follows: 
\begin{align}
\Sigma(\epsilon) \rightarrow \Sigma(\epsilon)\left(  h(-\epsilon -\Delta)  + h(\epsilon) \right).
\label{eq:filter}
\end{align}  
This kind of energy filtering is crude, but rather common in the literature.\cite{flage-larsen2012,Bahk2013:filtering} 

Figure~\ref{fig:Thermo_mu} shows calculated thermoelectric properties of ZnSb as a function of the Fermi level $\mu_{\rm F}$. The left side presents results at $T=~300~{\rm K}$, the right at $T=~500~{\rm K}$.
Panels \emph{a}) show the Hall carrier concentration,
 \emph{b}) the Seebeck coefficient, \emph{c}) the conductivity, and \emph{d}) the power factor. 
The full black curves show the constant relaxation time results for bulk ZnSb including the band gap correction specified above. 
The stark contrast with the dashed one, based on the bare PBE gap, 
underlines the importance of this correction.
With the low PBE gap, minority carrier contributions become significant for low and moderate doping, 
severely reducing the peak Seebeck value.
Further, the asymmetry of $\Sigma(\epsilon)$,
as seen in figure~\ref{fig:electonicStruc}, 
reflects a favoring of electron transport over hole carrier transport, resulting in a negative  Seebeck coefficient at Fermi levels close to the band edges. 
The asymmetry is also reflected in the shape of the conductivity and power factor, indicating that ZnSb could be a better n-type thermoelectric than a p-type,\cite{Bjerg20122111, Niedziolka2014,Jund201285} provided that stable n-type ZnSb with suitable doping concentration could be prepared.  So far no successful n-type has been reported while the difficulty has been rationalized by the easy formation of Zn vacancy type defects acting as acceptors.  In this paper, emphasis has thus been on the regular p-type variant.

The effect of various degrees of energy filtering 
  is shown with the thin green and dotted red curves in figure~\ref{fig:Thermo_mu}.
  Energy filtering drastically increases the peak Seebeck coefficient and power factor, but also shifts the peak positions to a lower Fermi level corresponding to higher p-doping concentrations. 
  The particularly high peak with an energy filtering parameter of $\Delta = 0.5~$eV can be linked to the shape of the band structure and to the density of states and transport spectral function in figure~\ref{fig:electonicStruc}. 
  At energies around $0.5~{\rm eV}$ additional bands start contributing causing a kink-like feature in these two functions. 

  Comparing the left and right subfigures, we find that for a given Fermi level, the Seebeck coefficient is lower at $500~\rm{K}$ than at $300~\rm{K}$, but as far as the power factor is concerned,  this is more than compensated by the increased conductivity, resulting in a higher value at $500~{\rm K}$.

  In the comparison with experimental data (in Sec.~\ref{sec:Comparison}), we will use the measured Hall carrier density at different temperatures $n_{\rm Hall}(T)$ to determine the Fermi level $\mu_{\rm F}(T)$. 
 We have then assumed that the holes and electrons scatter equally (but possibly depending on the energy of the band).
This is a minor approximation, since the transport properties are dominated by the majority carriers for the Hall carrier concentrations and temperatures considered here (when assuming the band gap is 0.56~eV).  
 In the case of constant scattering time, the Fermi level could thus be obtained for each temperature by solving the following equation:
  \begin{align}
    n_{\rm Hall}(T)r_H = \int_{-\infty}^\infty \diff \epsilon \, f_{\rm FD}(\epsilon-\mu_{\rm F}) \rho(\epsilon) \sign(\epsilon)  + N_{\rm val}.
    \label{eq:find_mu1}
  \end{align}
  Here $N_{\rm val}$ is the number of valence electrons in the system and $r_H$ is the Hall factor.
  For simple energy-dependent scattering (equation~(\ref{eq:tau})), we  used the Hall factor\cite{lundstrom2009fundamentals} $r_H(s) = \Gamma(2s + 5/2)\Gamma(5/2)/\left( \Gamma(s + 5/2)\right)^2 $ and related the Hall mobility to the drift mobility. 
  Here, $\Gamma$ is the gamma-function. For reference, $r_H(0) =1$,  $r_H(-0.5) \approx 1.18$, and $r_H(1) = 1.4$. This expression ignores non-parabolicity. This is in line with the use of simple scattering models also derived for parabolic bands. 

Care must be taken in determining the Fermi level
when energy filtering is included in the model, since filtered electrons do not contribute to the Hall carrier concentration.  
Thus, if a filter is used on $\rho(\epsilon)$ in equation~($\ref{eq:find_mu1}$),
the reference number of valence electrons $N_{\rm val}$ should be adjusted accordingly. 
Further, the Hall correction factor and simple energy-dependent relaxation time approximations 
become inappropriate as they are developed for parabolic bands. We have therefore only combined energy-filtering models with the constant relaxation-time approximation.

The thermoelectric transport properties were calculated using the
  \textsc{BoltzTraP}~\cite{boltztrap}  software package to 
  generate  the density of states $\rho_i(\epsilon)$ and the transport spectral functions $\Sigma_i(\epsilon)$ for each band $i$ at constant scattering time.
  Next, equations~(\ref{eq:thermo},\ref{eq:find_mu1}) were solved in a post-processing step using \textsc{scipy}\cite{scipy} routines in \textsc{python}.

  \begin{figure}
    \includegraphics[width=0.95\columnwidth]{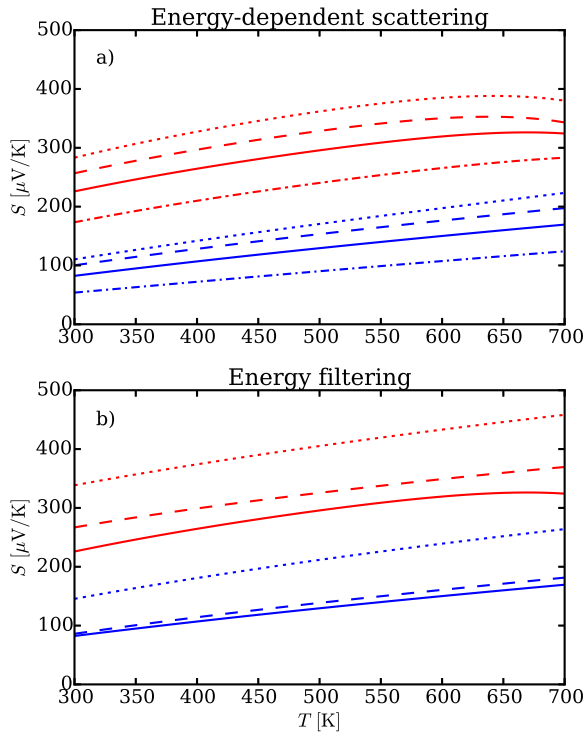}
-   \caption{Seebeck coefficients calculated using energy-dependent scattering (a) and energy filtering (b). The red and blue curves are for fixed Hall concentrations of $10^{19}$ and $10^{20}{\rm cm}^{-3}$, respectively. The solid curves are for a constant scattering time, $\tau_0 = 1.0\times 10^{-14}{\rm s}$. Energy-dependent scattering according to equation~(\ref{eq:tau}) is shown with an exponent of  $s=3/2$ (dotted curves), $s=1/2$ (dashed curves), and $s=-1/2$ (dash-dotted curves). In (b), energy filters (equation~(\ref{eq:filter})) of respectively 0.25~eV (dashed curves) and 0.5~eV (dotted curves) are introduced. \label{fig:SeebeckScatterOrfilter} }
  \end{figure}

  \subsection{Potential of energy filtering for ZnSb}
  \label{sec:potential}

Energy filtering greatly enhances the peak Seebeck coefficient of ZnSb, as shown in figure~\ref{fig:Thermo_mu}.
At the same time it severely reduces the electrical conductivity at a given Fermi level, since a significant number of charge carriers do not contribute to the transport anymore. 
However, the Fermi level may be manipulated if the doping level can be controlled. In that case, as the Fermi level approaches the filtered region, conductivity can be considerably increased, resulting in a strongly enhanced power factor. This is particularly so when filtering allows additional bands to contribute, as discussed above for the case of $\Delta=0.5$~eV.

Energy-dependent scattering can also enhance the Seebeck effect.
In fact, filtering can be viewed as an extremely energy-dependent form of scattering, as e.g.\ discussed by Bahk and coworkers.\cite{Bahk2013:filtering}  Whereas filtering may be appropriate as a crude model of the scattering or trapping caused by extended energy barriers such as grain boundaries,\cite{Nolas:Filtering,Narducci2015, Neophytou2013}
energy-dependent expressions are better suited to account for scattering by charged impurities such as acceptors or even charged nanoinclusions.\cite{Bahk2013:filtering}

  In figure~\ref{fig:SeebeckScatterOrfilter}, we compare the Seebeck coefficient as a function of temperature for different Hall carrier concentrations and different scattering/filtering accounts. 
  In the upper panel, we compare 
  the Seebeck coefficient for constant scattering time with energy-dependent scattering following equation~(\ref{eq:tau}) with $s=1/2$ and 3/2.  In the lower panel, we repeat the comparison for two different energy filtering parameters (equation~(\ref{eq:filter})) $\Delta = 0.25~$eV and 0.5~eV. 
  The figures illustrate how both energy-dependent scattering and filtering generally enhance the Seebeck coefficient. 
  The picture is somewhat more complex with energy filtering: the Seebeck coefficient is not always enhanced and the largest filtering parameter affects the results far more than the smallest. 
These effects arise because the Fermi level is shifted to keep the Hall carrier concentration fixed and multiple bands start contributing to the conduction for the largest filtering parameter.

  \begin{figure}[t!]
    \includegraphics[width=0.95\columnwidth]{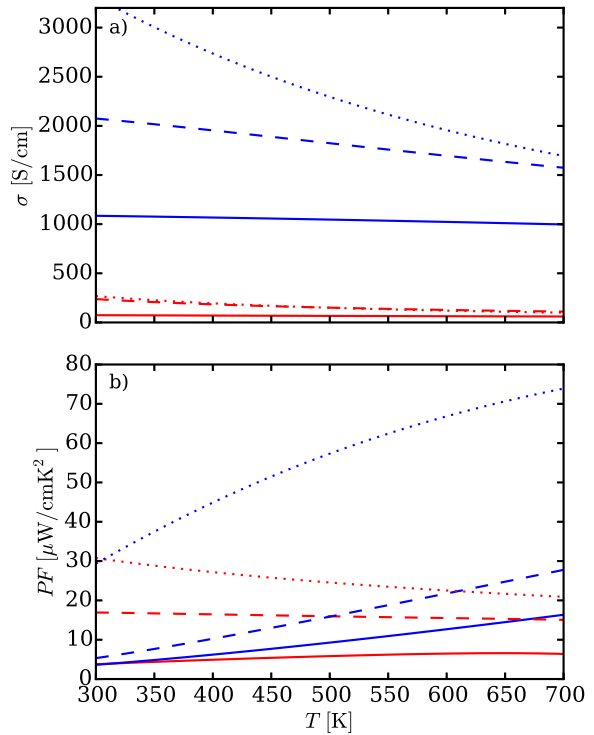}
    \caption{The effect of Hall concentration and energy filtering on the conductivity $\sigma$ (a) and power factor {\em PF} (b) in ZnSb. 
  The constant relaxation time is $\tau_0 = 10^{-14}{\rm s}$. 
  Following the conventions in figure~\ref{fig:SeebeckScatterOrfilter}(b), the red and blue curves are for fixed Hall concentrations of $10^{19}$ and $10^{20}{\rm cm}^{-3}$, respectively. The full curves are for a constant scattering time, while the dashed (dotted) curves have an energy filter of 0.25 (0.5) eV. The figures demonstrate how the power factor can be greatly enhanced with energy filtering. \label{fig:PF} }
  \end{figure}
 
The results of the energy filtering shown here are consistent with the data in figure~\ref{fig:Thermo_mu}. 
For instance, it is evident that decreasing the carrier concentrations (move to the right in figure~\ref{fig:Thermo_mu} a)) leads to increasing the Seebeck coefficient (move to the right in figure~\ref{fig:Thermo_mu} b)). 

  The slight dip in the Seebeck coefficient at $T=700~{\rm K}$ (red curve) for the lowest Hall carrier concentrations arises from minority carrier contributions. 
  When energy filtering is included this dip is absent; one effect of energy filtering is to increase the effective band gap by the same amount as the filtering parameter. 

  Figure~\ref{fig:PF} shows the conductivity and power factor as a function of temperature for the same filtering parameters and Hall carrier concentrations as in figure~\ref{fig:SeebeckScatterOrfilter}. In the upper panel, we find as expected that conductivity increases with the Hall carrier concentration. 
That filtering seems to enhance conductivity reflects that we have compared conductivities for different Hall carrier concentrations, only accounting for mobile holes and electrons.
  Depending on the physical mechanism causing filtering-like effect---instead of merely being passive, electron states could for instance also be removed from the active region---the effective doping concentration could dwarf the Hall carrier concentration. 
  Compare, for instance, the Hall carrier concentration curves with and without energy filtering in figure~\ref{fig:Thermo_mu}(a).
 With such high hole densities, the true potential profile in a sample with filtering barriers present could be strongly interconnected with the hole concentration.\cite{EnergyFiltering:GrainBoundaries}

The lower panel of figure~\ref{fig:PF} shows the corresponding power factors.
The crossing curves demonstrate that the optimal Hall carrier concentration for a given filtering parameter depends strongly on the target temperature.
Further, the optimal doping concentration for the Seebeck coefficient differs widely from the optimal one for the power factor (figure~\ref{fig:SeebeckScatterOrfilter}); for instance, at 700~K and a filtering parameter of $\Delta =0.5$~eV, the highest Hall carrier concentration considered ($2\times10^{20}{\rm cm}^{-3}$) results in both the lowest Seebeck coefficient and the highest power factor. Conversely the curve with the highest Seebeck coefficient corresponds to the lowest power factor. 
 
\begin{figure}
  \includegraphics[width=0.95\columnwidth]{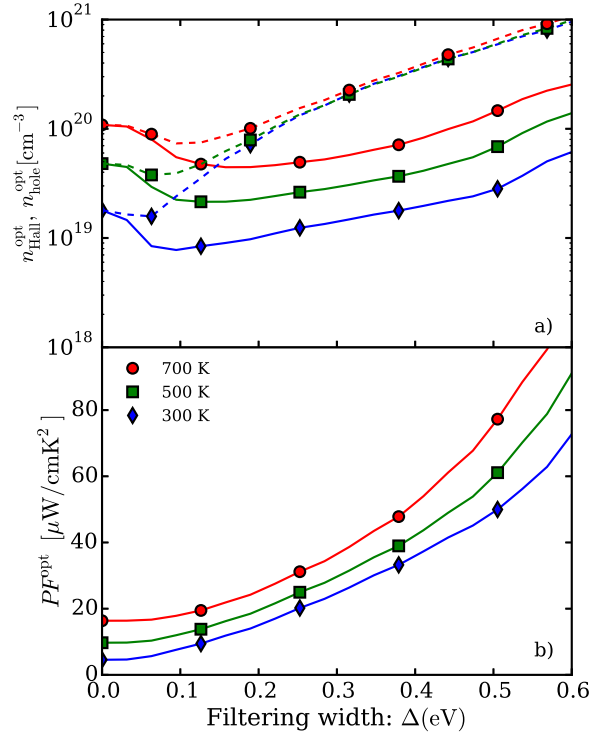}
  \caption{ The optimal charge carrier concentration (a) corresponding to the optimized power factor (b) for different filtering parameters $\Delta$.
  In (a), the solid and dotted curves represent the optimal Hall concentration and hole concentration for 300 (blue, diamonds), 500 (green, squares), and 700~K (red, circles). 
  It is shown in (b) how the optimal power factor increases with filtering parameter at the same temperatures as in the upper panel. 
    \label{fig:PowOpt} } 
  \end{figure}

Figure~\ref{fig:PowOpt} presents the optimal power factor and accompanying carrier concentration as function of the filtering width. 
In the lower panel, the optimal power factor is shown as a function of the filtering parameter $\Delta$.
A filtering parameter of 0.5~eV e.g.\ results in a tenfold increase in the power factor at 300~K. 
The relative enhancement is somewhat lower at higher temperature, but the power factor is nonetheless significantly higher than for lower temperatures. 
In the upper panel, the solid curves show the optimal Hall carrier concentration for the given filtering parameter and the dashed ones show the corresponding hole concentration (under the assumption that the filtering mechanism simply blocks propagation of filtered electrons). 
As the filtering parameter increases, the optimal hole concentration can easily become more than ten times larger than the Hall carrier concentration. 
Thus extremely high hole concentrations are required to optimize the power factor. This is the reason we have not evaluated filter widths beyond 0.5~eV, even if the power factor continues to increase as the filter width is increased further. At a certain point it is not realistic to obtain the carrier concentration required to optimize the power factor. We have somewhat arbitrarily selected 0.5~eV as the limit, since this would require an order of magnitude higher carrier concentration than the Hall concentration.
However, for small filtering parameters, the optimal carrier concentration might be slightly lower than without filtering. In this case, the enhancement of the Seebeck coefficient outweighs the reduction in the conductivity.  

  \section{Thermoelectric properties of bulk and nanostructured ZnSb}
  \label{sec:Comparison}
   \begin{figure}
    \includegraphics[width=0.95\columnwidth]{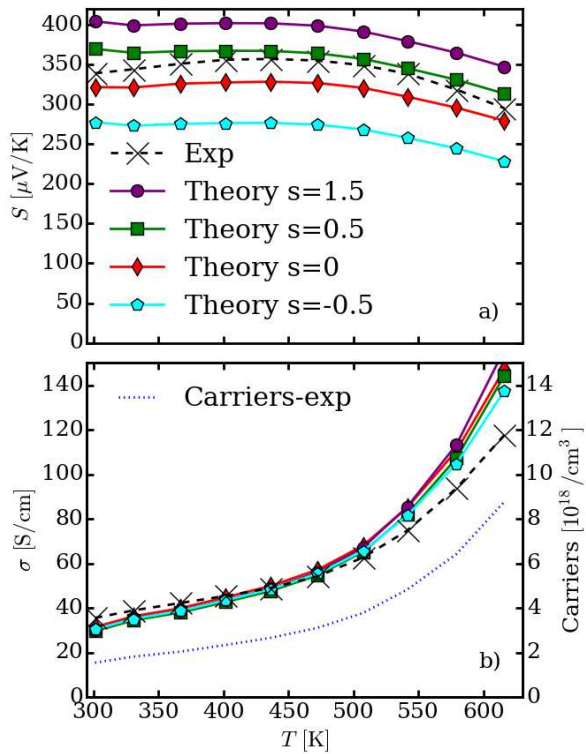}
    \caption{Seebeck coefficient $S$ (a), conductivity $\sigma$ (b), and experimental carrier concentration (dotted blue curve in (b), right axis) of a bulk ZnSb sample. 
    Crosses connected by black, dashed lines correspond to experimental data, while the filled symbols connected by solid lines correspond to calculated results based on the measured Hall carrier concentration using energy dependent scattering mechanisms according to equation~(\ref{eq:tau}) with $s=3/2$ (purple, circles), $1/2$ (green, squares), 0 (red, diamonds), and $-1/2$ (cyan, pentagons). \label{fig:Bulk} }   
  \end{figure}

  \subsection{Comparison with bulk reference sample}

  In comparing theory and experiment, we first considered a 
  nominally undoped bulk-like sample 
with a significant intrinsic carrier concentration. The grain size of this sample was measured to be $0.2~\mu$m by using the FWHM from the X-ray diffractogram. Note that the grain size distribution is also very important for thermoelectric properties; however, this was not available with our methods. In the calculations, we used the measured Hall carrier concentration to determine the Fermi level at each temperature, while the value of the constant relaxation time $\tau_0$ was subsequently obtained by fitting the temperature-dependent calculated electrical conductivity to the experimentally measured one. 

  In figure~\ref{fig:Bulk}, the upper panel compares the calculated Seebeck coefficient (full curves) with the measured one (dashed curve), while the lower panel compares the experimental conductivity with the calculated one, using the fitted relaxation time. 
  The constant relaxation time was used as a parameter to fit the calculated to experimental conductivity curves 
   in the temperature range between $300$ and $500$~K, and was then found to be $\tau_0=1.35\times10^{-14} {\rm s}$.
  The dotted curve shows the measured Hall carrier concentration (right axis).

  The reasonable agreement between theory and experiment for scattering parameters $s=0$ and 0.5 indicates that our relatively simple model based on full bands generated with DFT and with a constant-time scattering reproduces the experimental temperature-dependent conductivity and Seebeck coefficient quite well.
  The small discrepancies could arise partly from
the crude scattering account and partly from inaccurate band curvatures
obtained with the PBE functional, which could affect the effective mass and nonparabolicity. 
Finally, the Hall carrier concentration varies strongly as a function of temperature, and any error in this measurement would influence the theoretical predictions.  
The Seebeck coefficient as a function of temperature in figure~\ref{fig:Bulk} reaches a maximum value at around 450~K before decreasing. This is qualitatively different from the monotonously increasing one for fixed carrier concentration, shown above in figure~\ref{fig:SeebeckScatterOrfilter}. The difference can most easily be rationalized by the rapid increase in Hall carrier concentration that was used in calculating the Seebeck coefficient in figure~\ref{fig:Bulk}. The strong dependence of the Seebeck coefficient on the carrier concentration can e.g.\ be seen be comparing panels a) and b) in figure~\ref{fig:Thermo_mu}. It is worth noting that a turning point of $S$ like the one seen in figure~\ref{fig:Bulk} is often used to estimate the band gap, using the Goldsmid formula.\cite{Goldsmid1999} In our case the turning point can be explained solely by the strongly increasing majority carrier concentration as a function of temperature, illustrating one of the potential pitfalls when using the Goldsmid formula for band gap assessment.\cite{ Gibbs2015}

\subsection{Including filtering for reference carrier concentration}

\begin{figure}[t!]
\includegraphics[width=.95\columnwidth]{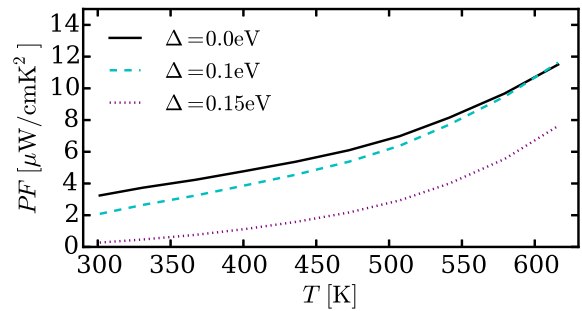}
\caption{Power factor as function of temperature for constant scattering time (black, solid curve) and filtering parameters $\Delta=0.1$~eV (blue, dashed curve) and $\Delta=0.15$~eV (purple, dotted curve).The hole concentration was fixed to that of the bulk sample. \label{fig:PowBulk}} 
\end{figure}
\begin{figure}[b!]
  \begin{center}
    {\includegraphics[width=7cm]{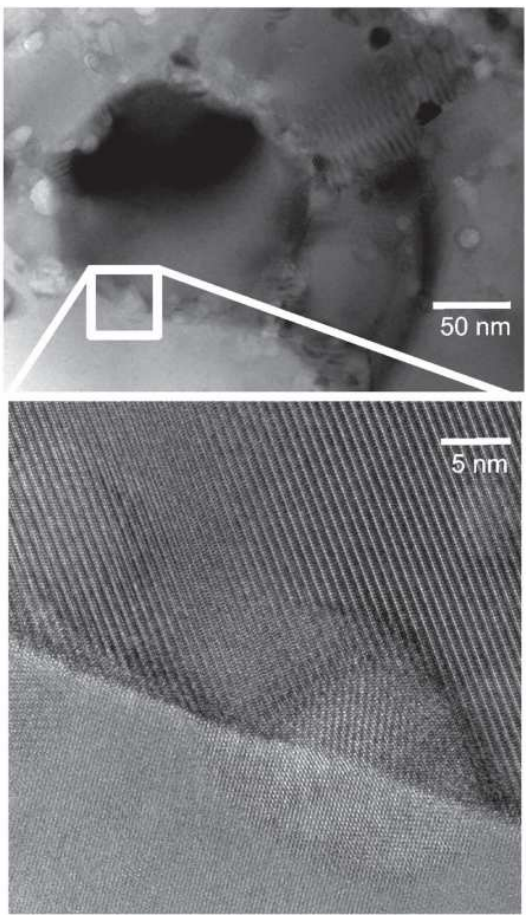}}
    \caption{Transmission electron microscopy (TEM) image of the nanostructured sample. The upper part depicts several grains of typical size, as well as a number of oxide precipitates. (Proven by electron diffraction on several different precipitates, not shown here.) The lower part has zoomed in on precipitates located along a grain boundary.}
    \label{tempicture}  
  \end{center}
\end{figure}

The carrier concentration we obtained from the nominally undoped bulk sample could be regarded as a typical one. But how would the performance be affected if we included energy filtering assuming that the hole concentration is kept fixed?
Figure~\ref{fig:PowBulk} shows that in this case the power factor is reduced as the filtering parameter $\Delta$ increases. 
This comparison differs inherently from that of figure~\ref{fig:PF}, where the power factor was calculated for different Hall carrier concentrations. This may be useful for comparing with experiment, but does not explore the effect of energy filtering for a given hole concentration.
The effective Hall carrier concentration may be significantly reduced by energy filtering, which is illustrated by the green and black curves in the upper panels of figure~\ref{fig:Thermo_mu}.
To achieve a high power factor, the hole concentration must be high enough to maintain a relatively high number of mobile carriers. 

\subsection{Comparison for nanostructured ZnSb}

\begin{figure}[t!]
\begin{center}
{\includegraphics[width=9cm]{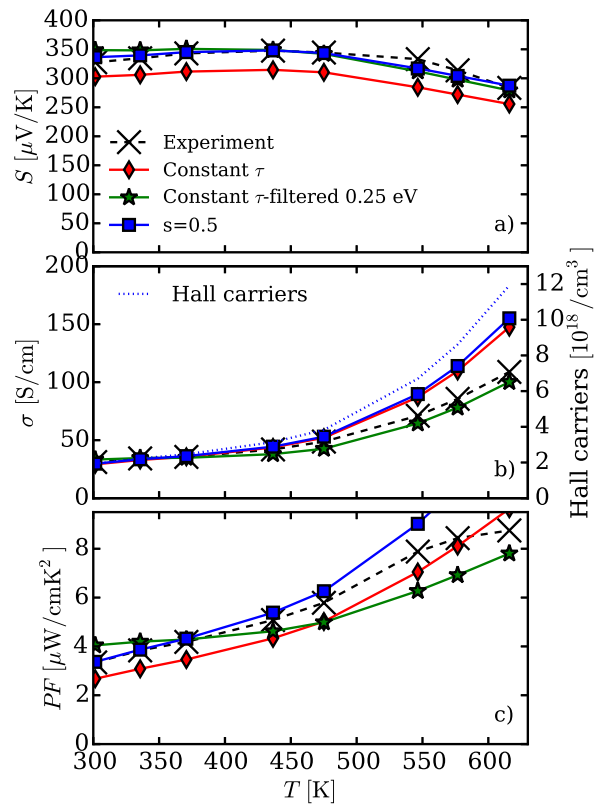}}
\caption{Thermoelectric properties of nanostructured ZnSb: Seebeck coefficient $S$ (a), electrical conductivity $\sigma$ (b), experimental carrier concentration (dotted blue curve in (b), right axis), and power factor {\em PF} (c).    Black crosses connected by dashed lines correspond to experimental data, while the filled symbols connected by solid lines correspond to calculated results based on the measured Hall carrier concentration using different scattering mechanisms: constant relaxation time (red, diamonds), constant relaxation time with a filter of $\Delta=0.25$~eV added (green, stars), and an energy dependent scattering according to equation~(\ref{eq:tau}) with $s=1/2$ (blue, squares). 
\label{fig:nano_thermo}  }
  \end{center}
\end{figure}

Having established the potential of energy filtering in ZnSb in Sec.\ \ref{sec:potential}, 
we now explore whether nanostructuring of ZnSb can be seen to induce energy filtering. The filtering mechanism could for instance be potential barriers at the grain boundaries, thus relying heavily on the grain size. To this end, we investigated experimentally the transport properties of two different ZnSb samples with average grain size of 70~nm (nanostructured) and $0.2~\mu$m (bulk), respectively. The processing of powders and pellet samples was briefly described in Sec.\ \ref{sec:experimental-methods} and in more detail in a previous paper.\cite{xins20152578}

Figure \ref{tempicture} shows a TEM image from the nanostructured ZnSb pellet, depicting 
a number of small grains as well as clustering of oxygen containing precipitates close to the grain boundaries. Such clusters could give rise to barriers hindering transport of low-energy charge carriers, making it a possible source of the filtering effect.
The mean grain size indicated by the XRD FWHM was 70~nm,\cite{xins20152578} consistent with the TEM image in figure \ref{tempicture}.

Transport properties of these nanostructured samples were then measured, and 
figure~\ref{fig:nano_thermo} shows a comparison between theory and those experiments. 
Three different scattering mechanisms are compared:  constant relaxation time; an energy dependent scattering (equation (\ref{eq:tau})) with $s=0.5$, corresponding to polar optical phonon scattering; and a combination of constant relaxation time with energy filtering (equation (\ref{eq:filter})) with $\Delta=0.25$~eV.
Like above, the Hall carrier concentration was used as input to determine the Fermi level at each temperature and scattering mechanism, 
followed by adjusting the relaxation time $\tau$ to fit the temperature dependent conductivity $\sigma$ to experiment in 
the temperature range  between 300 and 500~K. The Seebeck coefficient is independent of the specific relaxation time.

We first note that we can achieve a reasonable agreement between theory and experiment for all the scattering mechanisms in figure~\ref{fig:nano_thermo}. The constant $\tau$ and $s=0.5$ mechanisms yield too fast increase of $\sigma$ when $T\gt 500$~K. 
Also, constant $\tau$ yields a too low Seebeck coefficient for all temperatures when compared with experiment. 
The best fit is thus achieved with the combination of constant $\tau$ with energy filtering, using a filtering parameter of 0.25~eV. The constant $\tau$ was found to be slightly lower in the nanostructured sample ($10~{\rm fs}$) than that found for the bulk sample ($13.5~\rm{fs}$).

The model using constant $\tau$ with energy filtering exhibits a good match with the experimental curves of the Seebeck coefficient and electronic conductivity. However, the fit is not so good for the power factor. This is because of small deviations contributing in the same directions of both $S$ and $\sigma$ and being magnified for the product. Because of cancellation of errors both the constant $\tau$ and the $s=0.5$ mechanisms appear to give a better fit to the power factor.

This reflects that the difference in quality between the different models is not huge. Also, the deviation in the Seebeck coefficient from experiment of the nanostructured sample using the constant relaxation time model is similar to that of the bulk sample shown in figure~\ref{fig:Bulk}. This simply reflects that the two samples display quite similar carrier densities, since the Hall concentration is decisive for the Seebeck coefficient in this material. This was demonstrated by performing similar experiments with other bulk and nanostructured samples (not shown here); the quantitative success of the constant-time scattering model in bulk samples was highly dependent on the charge carrier concentration, and the Seebeck coefficient was quite similar in bulk and nanostructured samples at similar carrier concentration. Also, the power factor was not enhanced by nanostructuring.

Thus, no new scattering mechanism can be seen to appear when going from  bulk to  nanostructured samples. In other words, there is no need to involve energy filtering or more energy-dependent scattering resulting from grain refinement as part of the mechanisms explaining the transport properties of the nanostructured samples in this study.
 
It would be interesting to repeat the measurements with even smaller grains, preferably comparable in size to the energy relaxation length. This might be feasible, since the average particle size of the as-milled powder from the cryomill is $\sim 10$~nm.\cite{xins20152578} The energy relaxation length is not known for ZnSb. It is significantly larger than 10~nm in lightly doped bulk silicon (0.89 $\mu$m at 270~K with a charge carrier concentration of $\sim 10^{15}$~cm$^{-3}$),\cite{porter2012} but may be in the same order of magnitude in nanostructured, heavily doped systems.\cite{kim2012} To achieve such small grains would require a faster annealing technique than the rapid hot press used in the present study, and a close eye should be kept on grain growth by limiting the temperature used in the experiments. 

It may also be interesting to perform similar experiments with lower amount of precipitates clustered around the grain boundaries. Even if nanoinclusions may yield more predictable filtering barriers than grain boundaries,\cite{ Zhang2014} a system featuring only grain boundaries might give a more pure signal of filtering which is easier to interpret.

The current study relied on undoped ZnSb to simplify the analysis and focus on the effect of nanostructuring on the scattering properties. If one succeeds creating a sample displaying clear signs of filtering, the next important step would be to combine this with intentional doping. This is required to move towards the peak power factor as seen in figure\ \ref{fig:Thermo_mu}(d). It remains to see if any dopant has sufficient solubility in ZnSb to reach this regime.

\section{Conclusion}
We investigated the theoretical potential of energy filtering in the promising thermoelectric material ZnSb. It was shown to be considerable, with up to an order of magnitude increases in the power factor compared to bulk samples. This required a filtering parameter of 0.5~eV and high Hall carrier concentration. Our theoretical analysis also indicated that energy filtering would yield very high Seebeck coefficients at low Hall carrier concentrations.

The theoretical predictions were then tested against experiments on nanostructured ZnSb. The assumption was that nanostructuring could lead to energy filtering, enhancing thermoelectric properties by selectively hindering the conduction of low-energy charge carriers. 
Nanostructured ZnSb samples were processed by cryogenic milling of ZnSb into very fine powder and pressing pellets with a rapid hot press. They were nominally undoped, but still featured charge carrier concentrations in the order of $10^{18}-10^{19}$~cm$^{-3}$. 

The samples displayed a relatively large variation of the Hall concentration as function of temperature, which resulted in the Seebeck coefficient displaying a quite flat behavior. Thus, to obtain meaningful comparison between experiments and theoretical modeling, we adjusted the Fermi level of the calculations to reproduce experimental carrier concentrations for each temperature. Furthermore, the observed electrical conductivity at moderate temperatures ($300-500$~K) was used to calibrate the scattering parameters (constant scattering time $\tau$ and filtering parameter $\Delta$). With those parameters fixed, the measured Seebeck coefficient and the power factor served as benchmarks of the various scattering models, in the hope that distinct features of the different models could rule out or support any of them.

Reasonable correspondence with the experimental data was obtained when using any of the following scattering models: (i) constant scattering time, (ii) constant scattering time combined with a filtering with height 0.25~eV, and (iii) polar optical phonon scattering ($s=0.5$). The constant time combined with filtering (ii) exhibited a slightly better correspondence with experiment, but not enough to support the introduction of an extra adjustable parameter (the filtering height) in addition to a hypothetical physical mechanism.

Our conclusion is that an average grain size of around 70~nm is not small enough to obtain filtering with substantial effects on the scattering properties and power factor of ZnSb. Whether it is possible to obtain filtering in ZnSb, and whether a smaller grain size would render the effects of filtering observable are still open questions.

\section*{Acknowledgements}
We are grateful for enlightening discussions with Espen Flage-Larsen and for access to experimental facilities at California Institute of Technology via G. Jeff Snyder. We acknowledge the Research Council of Norway for financial support through the projects NanoThermo and Thelma. The computations were carried out using a grant from the Notur consortium.

\bibliography{library} 
\end{document}